\documentclass[12pt]{iopart}
\usepackage{graphicx,color,psfrag}
\usepackage{amssymb,subfig}

\begin{document}

\title{Tug-of-war in motility assay experiments}

\author{Daniel Hexner and Yariv Kafri}

\address{ Department of Physics, Technion-Israel Institute of Technology, 32000 Haifa, Israel}


\begin{abstract}
The dynamics of two groups of molecular motors pulling in opposite directions on a rigid filament is studied theoretically. 
To this end we first consider the behavior of one set of motors pulling in a single direction against an external force using a new mean-field approach. Based on these results we analyze a similar setup with  two sets of motors pulling in opposite directions in a tug-of-war in the presence of an external force. In both cases  we find  that the interplay of fluid friction and protein friction leads to a complex phase diagram where the force-velocity relations can exhibit regions of bistability and spontaneous symmetry breaking. 
Finally, motivated by recent work, we turn to  the case of motility assay experiments where motors bound to a surface push on a bundle of filaments. We find that, depending on the absence or the presence of a bistability in the force-velocity curve at zero force, the bundle  exhibits anomalous or biased  diffusion on long-time and large-length scales. 
 \end{abstract}



\maketitle

\section{Introduction}
Molecular motors are proteins which convert chemical energy into mechanical work. In many cases, relating to both \textit{in vivo} and \textit{in vitro} situations, they act together in large groups. Among the numerous examples are, myosin motors acting in muscles \cite{alberts}, kinesin motors pushing microtubules or myosin motors acting on actin filaments in motility assay experiments (see for example \cite{MotilityAssay1,MotilityAssay2}), and the extraction of membrane nanotubes by kinesin motors \cite{nanotubes,Tailleur,Ocampas1}. It is now well established that motors can exhibit a wide range of collective behaviors. Many times the collective behavior results in an oscillatory ``like'' motion  \cite {NCD,MuscFlagSim,Flag,SacroOsc,Electric} where the velocity changes abruptly between two distinct values \cite{NCD,JulicherSim}.

Frequently the setup is such that the motors all act together in a certain  direction. This occurs, for example, in motility assays where the filaments have a well defined polarity. In some cases, however, the picture is different and two groups of motors pull in opposite directions. This is the case, for example, in muscle contraction, active gels with myosin minifilaments \cite{MacK,Kruse,Swamy}, contractile ring that forms during cytokinesis, vesicles carried by both kinesin and dyenin along a microtubule bundles \cite{KinDyn} and more. Recently, such a scenario has also been realized in motility assay experiments. In one set of experiments \cite{KinNCD} a microtubule is acted upon by both NCD motors and homotetrameric kinesin-5 KLP61F in opposing directions. In another set, bundles of actin filament of opposing polarity are placed on a surface covered with myosin motors \cite{AnnBiDirectional}. It is very common in such experiments to observe an oscillating like behavior where a velocity, say of a bundle of filaments, changes between two distinct values.

 So far theoretical studies of two classes of motors acting in opposite directions in a tug-of-war have  focused on small groups of processive motors (which hardly detach from their track) acting on a fluid membrane \cite{TugOfWar}. 
In this paper we study theoretically a tug-of-war scenario for non-processive motors acting in large groups on a rigid filament. We focus on motility assay experiments as discussed above, although many of our results can be easily extended to other scenarios.  It is the central aim of this paper to analyze the different kinds of behavior that such a system can exhibit. To do this we generalize a discrete model, first introduced in \cite{vilfan}, to incorporate two groups of motors pulling on a bundle in opposite directions.  

To this end we first revisit the usual scenario where a rigid filament is pulled by a specified number of motors of one type. We analyze the model through a new mean-field approach which allows a straightforward derivation of velocity-force relations, where the force is exerted by some external agent on the filament. Our  mean-field approach gives rise to behaviors not observed in previous treatments of this system \cite{vilfan,JulicherReview,LeiblerHuse}. We find four distinct types of force-velocity curves shown in figures \ref{fig:phase1}(a) and \ref{fig:phase1}(b). In particular the force-velocity relation exhibits distinct bistable behaviors which result from the viscosity of the fluid and the protein friction. The bistability manifests itself  dynamically through an oscillatory like behavior where the velocity changes between two distinct values. Our analysis illustrates that there is a distinction between bistability which arises due to a ``fluid'' viscosity studied in \cite{JulicherReview} from that of  one which is caused by protein friction and studied in \cite{StickSlip}.  The model shows both. 

\begin{figure}
\centerline{\includegraphics[width=.5\textwidth]{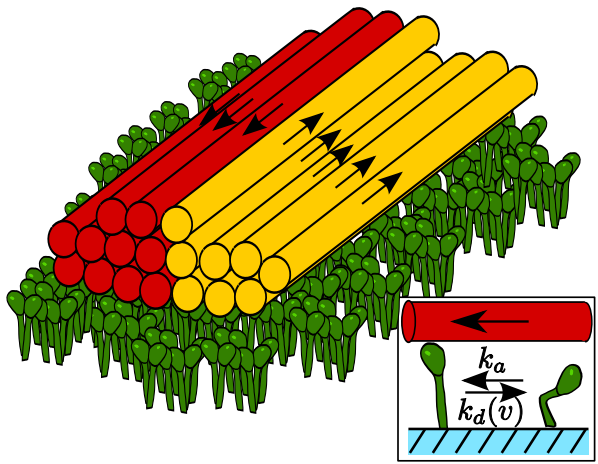}}
\caption{The setup we consider is of a bundle composed of many filaments set on a surface of motors. Note, that the analysis presented in the paper does not change if filaments of different polarity are mixed. In the inset a motor is shown with the transition rates between the attached and detached states.}
\label{fig:setup1}
\end{figure}

Building on these results we study the case of a bundle of filaments pulled in opposite directions by two groups of motors, with a specified number of each type, and acted upon by an external force (see \fref{fig:setup1} for a setup with no external force). Such a scenario could be realized using single molecule experiments. We find five different types of possible  force-velocity relations with as much as four regions of bistability (see figures \ref{fig:Phase2}(a) and \ref{fig:Phase2}(b)). Using existing data for myosin we discuss the force-velocity relation expected in a tug-of-war between two sets of myosin motors. We also present a systematic study of the dynamics of the system which result from the bistability. In the limit $N\rightarrow \infty$ the system can exhibit spontaneous symmetry breaking. 

We conclude by considering motility assay experiments. In contrast to the scenario discussed above the number of motors acting in each direction now fluctuates as a function of the location of the bundle on the substrate. Different behaviors are found if the system is in a bistable regime or not. When the system is not in a bistable regime, as the size of the bundle increases, its motion becomes irregular (see  \fref{fig:typxt1}). The bundle gets trapped for very long times at specific locations and in a large bundle limit (defined carefully in the text) the mean square displacement of the bundle grows as $log^4(t/ \tau_0)$. Here $t$ denotes time and $\tau_0$ sets a time scale. Moreover, at locations where the bundle remains trapped it displays an oscillatory like behavior despite the fact that there is no bistability. When the force-velocity relation is bistable the motion is also irregular. However, now in the large bundle limit the mean square displacement of the bundle is linear in time (diffusive) and the bundle is expected to have a mean, non-zero, velocity. Our results provide an explanation for the experiments of \cite{AnnBiDirectional} without resorting to their assumption of cooperative unbinding of the motors. 

The structure of the paper is as follows. In \sref{model} we define the model. In \sref{meanfield}  the model is analyzed using a mean-field approximation. In \sref{dynamics} the dynamics of the motion are examined. \Sref{disorder}  studies the case where the number of motors varies randomly in space. Finally we summarize in \sref{conclusions}.
\section{The model\label{model}}
The scenario of interest is of two sets of motors pulling in
opposite directions on a common backbone - a tug-of-war. One set pulls
in the ``plus'' direction with a force $F_{+}$ while the other pulls with a
force $F_{-}$ in the ``minus'' direction (see \fref{fig:setup1}). The motion at low Reynolds number is governed by the force balance
equation where the force on the filament is countered by the viscous
drag of the filament, namely
\begin{equation}
\overline{\eta}v=F_{+}+F_{-}+F_{ext}.\label{eq:ForceBalance}
\end{equation}
 On the left hand side of the equation $\overline{\eta}$ is the viscosity and $v$ is the velocity of the filament. 
 On the right hand side are the forces acting on the filament and
we include a possible contribution from an external force, $F_{ext}$.  We assume a completely rigid filament. To find $F_{\pm}$ a microscopic model of the motors is needed.

The model we use  is based on the one introduced in \cite{vilfan}. Illustrated
in \fref{fig:setup1}, it consists of two motor states denoted by $a$ and $d$. In
the $a$ state the motor is attached to the filament, while in the $d$
state it is detached. The transition rate between states $d$ and
$a$ ($a$ and $d$) is denoted by $k_{a}$ ($k_{d}$). As shown
in \fref{fig:setup1}, we assume that when the motor binds to the filament it enters
a tense state, where some ``spring-like'' degree of freedom, $x$,
is extended from an equilibrium position $x=0$. The motor then exerts
a force as the spring relaxes and finally completes the cycle by
detaching from the filament. In general we expect $k_{d}$ to increase
with the tension on the motor. Following \cite{vilfan}, we take $k_{d}=\omega_{d}exp\left(\alpha\left|x\right|\right)$
which is consistent with the usual Kramers form of rates with $\alpha=Kl/k_{B}T$.
Here $K$ is a spring constant, $l$ is a microscopic length, $T$
is the temperature and $k_{B}$ is the Boltzmann constant. In addition
to being physically motivated, as stressed in \cite{vilfan}, simpler forms,
for example monotonically decreasing with $x$, do not yield an oscillating like 
behavior which is the focus here. 

To analyze the model we employ  a mean-field approximation. This  is done by  writing  self-consistent equations for the force generated
by the motors and the transition rates. To this end we relate the displacement 
$x$ (of the ``spring-like'' degree of freedom) to the velocity of the filament, $v$. We expect the approach to hold
for a large number of motors, when the fluctuations in the velocity
are negligible. The self-consistent equations are obtained as follows.
We denote the velocity of the filament by $v$, the time since the
motor attached to the filament by $t$ and the initial extension of
the spring after attaching to the filament by $x_{0}$. The rate $k_d(x)$ can be expressed using $x_0$, $v$ and $t$, namely $k_{d}\left(x_{0},v,t\right) =k_d(x)=k_d(x_0-vt)$. The probability density of detaching
at time $t$ is then $p\left(t,v,x_{0}\right)=exp\left(-\int_{0}^{t}k_{d}\left(x_{0},v,t'\right)dt'\right)k_{d}\left(x_{0},v,t\right)$
so that a self-consistent detachment rate, $k_d(v)$, can be defined
through:
\begin{equation}
 \frac{1}{k_d\left(v\right)}=\int dx_{0}q\left(x_{0}\right)\int_{0}^{t}dt'p\left(x_{0},v,t'\right)t'.
 \end{equation}
Here $q(x_{0})$ is the probability density of the springs attaching
to the filament with an extension $x_{0}$. Note that within our approximation we replace the several time scales present in $p(t,v,x_0)$ by a single one. The self-consistent rates
can be easily obtained for different choices of $q(x_{0})$ numerically.

We find that for generic choices $1/k_d(v)$ is peaked around some value
$v=v_{0}$ and decays monotonically to zero on both sides of the peak. 
Fast velocities (positive or negative) imply a fast growth of $k_d(x)$. This in turn makes $k_d(v)$  larger. The peak around the finite value of $v_0$ results from the finite positive average of $q(x_0)$ which is assumed. 
 Moreover, we find that the exact form of the function is unimportant for the qualitative results expected from such a simple model \cite{unpublished} (this will become evident later). 
To this end, we use a simplified analytical form, $k_{d}\left(v\right)=\kappa\left(\left(v-v_{0}\right)^{2}/W+1\right)$,
which captures all the important features of the exact form derived
by the procedure above. Here $\kappa$ is the binding rate at $v=v_{0}$
and $W$ is a scale parameter with units of velocity squared. A similar
procedure is used to define $k_{a}$. Since in the detached state
the spring is not stretched, it is easy to see that the above procedure
leads to an attachment rate which is independent of $v$ (see also
\cite{vilfan,LeiblerHuse}).

Next, we need to specify the force exerted by the
motor. This force, of course, varies with time. Within our approach
 we replace the time dependent force by its average,
$\langle f(v)\rangle$, over the attachment time to the filament $1/k_{d}(v)$.
While this can be done formally it is easy to see that to linear order
in $v$, $\langle f(v)\rangle=G-\gamma v$. Positive velocities tend
to initially release the tension in the spring decreasing the force.
The term $\gamma v$ is the leading order behavior of the protein
friction \cite{LeiblerHuse} which arises due to the energy dissipated
when a tense motor unbinds. To see this,
consider motors that are unable to actively generate force, so that $G=0$. Clearly,  $-\gamma v $ is the force resulting from the elastic element of the bound motors being stretched by an external force.  In general $\gamma$ can have a non trivial, details sensitive, dependence on the velocity. However for large velocities
motors are quickly detached from the filament and the effect of the
friction becomes less important. This will be shown in the treatment
of the collective behavior of the motors presented below. 

Before turning to the two sets of motors problem it will be useful to first analyze the single set problem, $F_-=0$,  as our solution to the two set problem relies on it.  \\

\section{Mean-field analysis\label{meanfield}} 
\subsection{One set of motors.}
In this section we will examine the case where a single set of motors operates
against an external force, namely $F_{-}=0$. In principle, since
the model constitutes a one-step process over the number of  motors in the $a$ state,  $N_{a}$,
a formal solution may be written for the steady state. However, it
is more instructive to examine a mean-field solution that is easily
generalized to two sets of motors. Based on the rates defined above
it is straightforward to write mean-field equations for $P_{i}=N_{i}/N$, 
the fraction of motors in each state
\begin{equation}
\partial_{t}P_{a}=P_{d}k_{a}-k_{d}(v)P_{a}.
\end{equation}
Here $N_{i}$ is the number of motors in state $i=d,a$ and $N$ is the
total number of motors. Note that a single motor is coupled to the rest
of the motors through the rate $k_{d}(v)$ which has a non-trivial dependence
on the velocity, $v$. The mean-field approximation, on top of the approximations described above, neglects correlations
between $k_{d}(v)$ and $P_{a}$. Furthermore, since the motors are identical
the same equation holds for all motors.

The stationary solution of these equations is easily solved and along
with the normalization condition $P_{d}+P_{a}=1$ yields 
\begin{equation}
P_{a}\left(v\right)=\left(1+\frac{k_{d}\left(v\right)}{k_{a}}\right)^{-1}.\label{eq:PofV}
\end{equation}
Note that the expression depends on $v$, the velocity of the filament.
To obtain $v$ as a function of, say $f_{ext}=F_{ext}/N$, one then
uses the solution self-consistently in the force balance equation,
setting $F_{+}=N_{a}\left\langle f\left(v\right)\right\rangle $.
To a first approximation we take $\overline{\eta}$ proportional to
the length of the filament. Assuming the motors are evenly spaced
out the overall viscosity can be expressed using the total number
of motors $N$ and $\eta$, the viscosity per unit distance between
motors, such that $\overline{\eta}=N\eta$. This results in:
\begin{equation}
f_{ext}\left(v\right)  =  \eta v+\left(\gamma v-G\right)P_{a}(v),\label{eq:FofV}
\end{equation}
with $P_a(v)$ given in  \eref{eq:PofV}. The right hand side of the equation is a sum of three terms. The
first, due to the viscosity of the filament, is monotonic in $v$. The
second results from the protein friction and the third from the force
exerted by the motors. The last two terms multiply $P_{a}(v)$ which
due to the functional form of $k_{d}(v)$ is a non-monotonic function
of $v$. As we show this implies that in some ranges of the parameters there is a region where for every value
of $f_{ext}$ there are three solutions for $f_{ext}\left(v\right)$.
Following \cite{JulicherSim} we take this as evidence for bistability. In fact, we demonstrate 
below that the competition between the three terms, with two non-monotonic,
can lead to a rather rich behavior with four distinct phases. For now we focus on the steady state solutions. Later on we analyze the dynamics which arise in the bistable regimes. \\

\subsection{Phase diagram}
\begin{figure}
\centerline{\includegraphics[width=.6\textwidth]{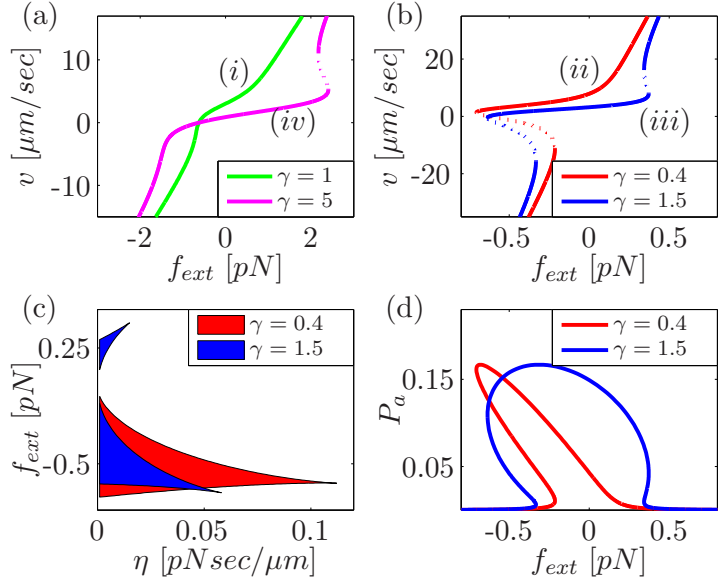}}

\caption{Results for a single set of motors. (a) force-velocity curve of types (i) and (iii).  Here $\eta=0.1 pN sec/\mu m$. (b) Force-velocity  curves of types (ii) and (iv). Here $\eta=0.01 pN sec/\mu m$. (c) Bistable ranges of $f_{ext}$ as a function of $\eta$ (shown in blue and red). (d) Fraction of attached motors with parameters as in (b). In (a) and (b) the dotted line signifies the regions of the solutions where $\partial_v f_{ext}(v)<0$.  In all figures $k_a/\kappa=1/5,~G=5 pN,~W=10 (\mu m/sec)^2,~v_0=2 \mu m/sec$. $\gamma$ is given in the graph in units of $pN sec/\mu m$ } 
\label{fig:phase1}
\end{figure}

In  this section we classify the different possible
force-velocity curves. Asymptotically, for large $f_{ext}$ most
of the motors are detached and the velocity is roughly $v\simeq f_{ext}/\eta$.
For smaller forces, depending on the parameter values, the contribution
from the non-monotonic terms can be important.  We find
four distinct regimes which qualitatively depend on $\gamma$, $G$ and $\eta$ as follows:

\noindent \textit{$(i)$~Large $\eta$, small $\gamma$ and small $G$ -
No bistable regime}: Here the first term in \eref{eq:FofV} dominates
due to the large viscosity so that the velocity changes monotonically
with the external force. The stall force, defined as $v\left(f_{ext}\right)=0$
occurs for $f_{ext}<0$, or in other words the velocity is positive
when there is no external force acting on the motors (see  \fref{fig:phase1}(a)). 

\noindent \textit{$(ii)$~Small $\eta$, small $\gamma$ and large $G$ -
Single bistable regime at $f_{ext}<0$}: Here the third term, due to the force
exerted by the motors, is large enough so that its non-monotonic behavior
becomes important. This is seen in  \fref{fig:phase1}(b) where the force-velocity curve
shows a region with three possible values of $v$ for a given value of $f_{ext}$. In particular there
is a region where $\partial_{v}f_{ext}\left(v\right)<0$ for one solution
and $\partial_{v}f_{ext}\left(v\right)>0$ for the two other solutions.
Since $\partial_{v}f_{ext}\left(v\right)<0$ implies a negative
mobility this suggests that the solution is unstable. The difference
between the two stable solutions is manifested through the number of motors in
state $a$. The solution with the larger velocity has more motors
in the attached state enabling it to counter the external force. The
solution with the smaller velocity has most of the motors detached,
resulting in a negative velocity. This can be seen by setting $\gamma=0$
in  \eref{eq:FofV} and is illustrated in \fref{fig:phase1}(d)
where we plot $P_{a}\left(f_{ext}\right)$.

\noindent \textit{$(iii)$~Small $G$, small $\eta$ and large $\gamma$ -
Single bistable regime at $f_{ext}>0$}: Here only the second non-monotonic
contribution due to protein friction induces a bistable regime.
This occurs as long as $\eta$ is not too small that the non-monotonic
behavior due to the force exerted by the motors becomes unimportant.
Now the force-velocity curve has a single bistable region located at
 $f_{ext}>0$ while  the velocity varies continuously for $f_{ext}<0$ (see  \fref{fig:phase1}(a)). 
 Unlike $(ii)$ both stable solutions have a positive velocity.

\noindent \textit{$(iv)$~Small $\eta$, large $\gamma$ and large $G$ - Two
bistable regimes}: Here both non-monotonic terms, due to the
protein friction and the force exerted by the motors become important. Interestingly, this can lead
to two distinct regions of bistability. In contrast to the   bistable  region
due to the force exerted by the motors ($ii$), in the additional bistable region both of the stable solutions
have a positive velocity. This is illustrated in  \fref{fig:phase1}(b)
where it is seen that increasing $\gamma$ adds a second bistable regime located 
at positive forces.

 For the specific choice we make here for $k_d(v)$ the results can be understood as follows. Due to the symmetric choice of the rate $k_d(v)$, the fraction of attached motors has the property $P_a(v+v_0)=P_a(-v+v_0)$.
Therefore, the force of the motors can be expressed in terms of symmetric and anti-symmetric functions relative to $v_0$, namely
\begin{equation}
  (G-\gamma v) P_a(v)= P_a(v) (G-v_0 \gamma) -\gamma P_a(v)(v-v0).
\end{equation}
The first term on the right hand side is symmetric with a single extrema and therefore  contributes a single non-monotonic region. The second  term on the right hand side is anti-symmetric and has two extrema located symmetrically around $v_0$ contributing  two bistable regions. Note that when $G=v_0 \gamma$ there is only an antisymmetric contribution. We stress that while this argument is specific for our choice of $k_d(v)$ the general structure is unchanged for other choices of non-symmetric $k_d(v)$. 

The different parameters can in principle be controlled, to some extent,
experimentally to observe possible transitions between the different
regimes. For example, $\eta$ can be controlled by the density of
motors along the filament and $\gamma$ can be controlled to some
extent by the $ATP$ concentration. It is well known that increasing $ATP$ concentration increases the unbinding rate \cite{Howard}. 
Therefore large values of $\gamma$ would correspond to small unbinding rates of the motors and hence small
$ATP$ concentration. $G$ might be tuned by changing the neck region of
the motor.
In  \fref{fig:phase1}(c) we illustrate two transitions
controlled by the viscosity $\eta$ --- one from two bistable regimes
to a non-bistable regime and one from a single bistable regime
to a non-bistable regime. The filled areas represent ranges of
$f_{ext}$ for which there is  bistable behavior for the first
transition (blue) and the second transition (red). For every value
of $\eta$, the region is calculated from the two adjacent extrema
of the mean-field $f_{ext}\left(v\right)$. For large enough viscosity
$\eta$, the velocity $v$ changes continuously with the external
force, $f_{ext}$. As $\eta$ is decreased beyond a threshold value
a bistable region emerges and the range of force values it encompasses
grows.

\begin{figure}
\centerline{\includegraphics[width=.5\textwidth]{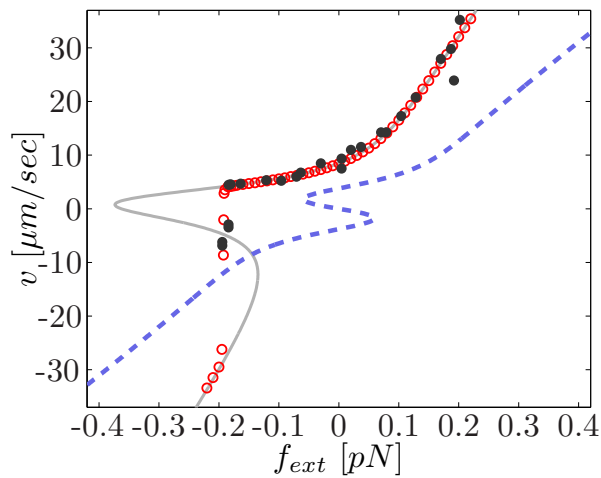}}
\caption{A fit of the theory to experimental data. Solid dots ($\bullet$) are data extracted  from reference \cite{Electric}. The solid line is the results of the mean-field theory and empty circles ($\circ$) represent the average velocity obtained from the simulations of the model using the fitted parameters of the mean-field results. The dashed line represents the predictions for tug-of-war between two equally sized groups of myosin-motors.} 
\label{fig:fit} 
\end{figure}

It is interesting that the simple model presented above accounts
well for  the  measured force-velocity curves
for myosin II. In  \fref{fig:fit} we show a fit of the mean-field
solution to the data of reference \cite{Electric}  along with numerical simulations
of the model. In contrast to the  mean-field solution, the experimental
and numerical data show that the positive velocity branch is
stable up to a certain force. In other words, up to this force the
motion of the filament with negative velocity is hardly observed.
We shall return to this point. It results from of the dynamical aspects
which are not captured by the mean-field solution that will be
discussed in a separate section and appears in our numerical simulations of the model (shown in the figure). To fit the data we used $\kappa /k_a =1/10$, $G = 4.5 pN$, $v_0=1.5 \mu m/sec$, $\gamma = 0.28 pN sec/\mu m$, $W=16 (\mu m/sec)^2$  and $\eta = 0.006  pN sec/\mu m$, which are in the range of values measured
independently  in references \cite{Howard} and \cite{fit} (see also references within). The experiments were performed using
an external voltage and we rely on their conversion between voltage and
force per motor. While in \cite{Electric} the points with negative velocity
are quoted to be stable our model predicts them to result from the motion
 of the filament in two opposite directions. Clearly more
data is needed in this region to improve the fit. Moreover the fit can be improved by using more elaborate forms for $k_d(v)$. 

The above analysis interpolated between two
regimes which were previously studied. In one class of models a joint rod and 
protein friction term was accounted for in the form of the first term on the right-hand side of \eref{eq:FofV} \cite{JulicherReview}. In another only protein friction in the form of the second term on the right-hand side of \eref{eq:FofV} was considered \cite{vilfan}.
We show that both lead to distinct bistable regimes
which can lead to a richer behavior than previously discussed. Our
focus in the paper, however, is a tug of war scenario. As we
illustrate, using the approach developed above, the analysis
becomes straightforward.
\subsection{Two sets of  motor  engaged in a tug of war}
We now turn to discuss the possible force-velocity curves for two
sets of motors pulling one against another in a tug-of-war.  Again we focus on the steady-state mean-field predictions and later  discuss dynamical aspects.
 The setup we consider is as follows. Two (or several) connected actin filaments of opposite
polarity form a bundle which is  set on a substrate of motors. In this section we assume
that the number of motors which operate on the bundle  is constant.
Thus, the filament (or filaments)  with positive polarity is acted upon by $N^{+}$ motors
pulling in the positive direction while the filament (or filaments) with negative
polarity is acted upon by $N^{-}$ motors pulling in the negative
direction (see  \fref{fig:setup1}).

Motivated by the experiments of reference \cite{AnnBiDirectional}, and for simplicity, we assume that both sets of motors, pulling in opposite directions,
are identical. The definition of the rates of the model then follows
as above but now with a modified force balance equation which accounts
for both sets of motors. Defining $N_{a}^{+}$ ($N_{a}^{-}$) to be
the number of attached motors pulling in the positive (negative) direction
the force balance equation now reads:
\begin{equation}
\eta v\left(N^{+}+N^{-}\right)=N_{a}^{+}\left(G-\gamma v\right)-N_{a}^{-}\left(G+\gamma v \right)+\overline{F_{ext}}.\label{eq:FV2sets}\end{equation}
Here $N^{+}+N^{-}$ is the total number of motors, $N_{a}^{+}\left(G-\gamma v\right)$
is the force produced by the motors pulling in the positive direction
and $-N_{a}^{-}\left(G+\gamma v\right)$ is the force produced by
the motors pulling in the negative direction. Finally $\overline{F_{ext}}$
is an external force (the overline is used to distinguish this  from the case of a single set of motors). As in the previous section the total viscosity
is taken to be proportional to the total number of motors.

Again we treat these equations within a mean-field approach replacing
$N_{a}^{\pm}$ by $N^{\pm}P_{a}^{\pm}$. Here $P_{a}^{+}$ ($P_{a}^{-}$)
is the fraction of motors in the attached state pulling
in the positive (negative) direction. Clearly, these satisfy  \eref{eq:PofV}
so that $P_{a}^{+}\left(v\right)$ and $P_{a}^{-}\left(v\right)$
are related through $P_{a}^{+}\left(v\right)=P_{a}^{-}\left(-v\right)$.
Using this with the expression for $f_{ext}(v)$ defined in  \eref{eq:FofV}
the force-velocity relation takes the form:
\begin{equation}
\overline{F_{ext}\left(v\right)}=N^{+}f_{ext}\left(v\right)-N^{-}f_{ext}\left(-v\right).\label{eq:2FVR}\end{equation}
Note that the relation only relies on the anti-symmetric nature of
the force exerted by the motors and additivity of the viscous terms.
In the following we consider the case $N^{+}=N^{-}\equiv N$, for
which $\overline{F_{ext}\left(v=0\right)}=0$. For this case the force-velocity curves, in terms of the normalized external force $\overline{f_{ext}(v)}=\overline{F_{ext}(v)/N}$,  are independent $N$. The force-velocity curves can be easily obtained using  \eref{eq:2FVR} by creating antisymmetric  combinations of the curves obtained for one set of motors. \\
\subsection{Phase diagram}
\begin{figure}
\centerline{\includegraphics[width=.6\textwidth]{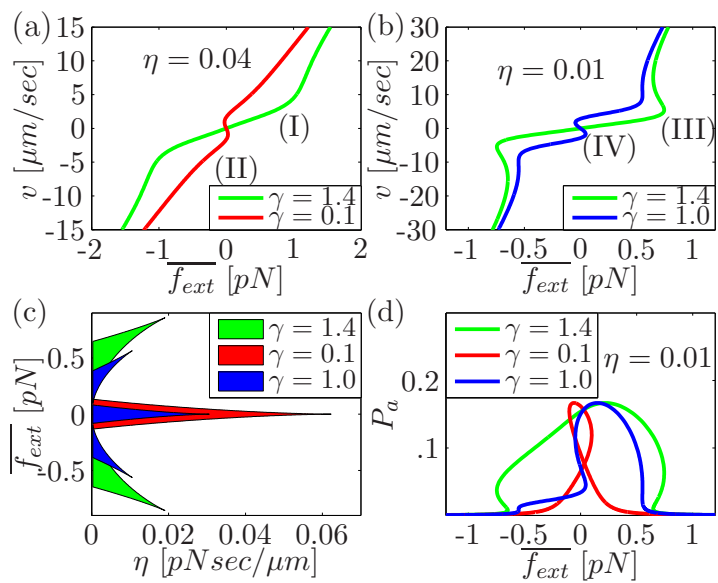}}
\caption{Results for two equally sized sets of motors pulling in a tug of war. (a) Force-velocity curves of types (I) and (II). Here $\eta =0.01$. (b) force-velocity curve  of types (III) and (IV). Here  $\eta =0.04$. (c) The range of values of the external force where there is a bistability as a function of $\eta$ (marked by green, blue and red areas). (d) The fraction of  attached motors pulling in the ``positive'' direction as a function of external force,  $P^{+}(\overline{f_{ext}})$. In all figures $k_a/\kappa=1/5,~G=5 pN,~W=10 (\mu m/sec)^2$.  $\gamma$ is given in the graph in units of $pN sec/\mu m$.  $\eta$ is specified in the graph in units of $pN sec/\mu m$. For the  green curves (I),(III)  $v_0=1 \mu m /sec$, for the red  curves (II)  $v_0 = 0.6\mu m /sec$ and for the  blue (IV)  curves $v_0=4 \mu m /sec$.}
\label{fig:Phase2}
\end{figure}

By following the above procedure we find four distinct generic structures for the force-velocity curve. Similar to the one motor case these  arise from an interplay of the various viscous terms and the force exerted by the motors. In contrast to the single motor case the generic curves cannot be easily classified according to the values of the viscosities and the force exerted by the motors.\\
$(I)$~\noindent \textit{Monotonic force-velocity curve:}
Here the force-velocity curve does not have any bistable regions (see  \fref{fig:Phase2}(a)). Such behavior can arise under several situations. The simplest one involves an antisymmetric combination of a monotonic force-velocity curves for a single motor (regime $(i)$ of the previous section). However under certain conditions it might also occur by antisymmetric combinations of any of the other regimes. This will occur when the non monotonic regime of one curve is weak enough so that when a monotonic contribution is added to it, it becomes monotonic. In all these cases, in the absence of an external force the average velocity of the rod is zero.\\
$(II)$~\noindent \textit{Single oscillating regime, centered around $\overline{f_{ext}}=0$:}
This can occur only when we combine single motor curves from regime $(ii)$ or regime $(iv)$ of the previous section. The latter combination requires that the non monotonic behavior due to the  protein friction is canceled by the antisymmetric combination. In this case at zero external force the system exhibits two stable solutions with velocities equal in magnitude but with opposite sign (see \fref{fig:Phase2}(a)). For the solution with positive velocity $N_{a}^{+}>N_{a}^{-}$, while for solutions with negative velocity $N_{a}^{+}<N_{a}^{-}$ (see \fref{fig:Phase2}(d)). Note that similar force-velocity curves have also been shown to exist theoretically for completely symmetric motors \cite{JulicherSim}.\\
$(III)$~\noindent \textit{Two non-monotonic regimes}
This occurs when single motor force-velocity curves from regime $(ii)$,  regime $(iii)$ or regime $(iv)$ of the previous section are combined. Now the force-velocity curve has two symmetric bistable regions at positive and negative forces (see  \fref{fig:Phase2}(b)). In the absence of any external force there is a single solution with zero velocity. In this case the fraction of attached motors $P_{a}^{\pm}$ is equal and relatively large.\\
$(IV)$~\noindent \textit{Three bistable regimes:}
This can occur only by combining force-velocity curves from regime $(iv)$ of the previous section. Now there is one bistable regime around zero external force and two antisymmetric bistable regimes at positive and negative external forces.\\
$(V)$~\noindent \textit{Four bistable regimes:} As evident from the above constructions it is also possible to find regimes where there are four bistable regimes by suitable combinations of curves for one set of motors. Two are at small positive and negative values of the forces and two are  at large ones. In fact, in some cases, the two small force regimes may be located one on top of the other. This leads to a regime around zero force with four stable velocities. By explicitly plotting the curves we find that both of these require a careful fine tuning of parameters. We therefore do not expect them to appear under generic conditions.

The transitions into the bistable regimes as a function of viscosity are illustrated in \fref{fig:Phase2}(c) for various parameters. Again we see that increasing the viscosity smooths out the force-velocity curve until it becomes completely monotonic.  In addition $P_{a}^{+}(\overline{f_{ext}})$ is seen to have a complex structure, as shown in \fref{fig:Phase2}(d), with the same number of bistable regions as the force-velocity curve.

In the case of two equally sized groups of motors, a simple criterion for bistability at $\overline{f_{ext}}=0$ can be found from the results for a force-velocity curve of a single set of motors. This is done  by requiring $\partial_{v}\overline{f_{ext}}|_{v=0}<0$, which can be rewritten using  \eref{eq:2FVR}:
\begin{equation}
\partial_{v}\overline{f_{ext}}|_{v=0} = 2 \partial_v f_{ext}(v)|_{v=0}<0.
\end{equation}
 Namely, the slope of the single motor set  force-velocity curve at stall force has to be negative. On a single set of motors this could be measured from  the slope of the force in constant velocity experiments \cite{StickSlip} around zero velocity. Constant velocity experiments are required as the zero velocity solution might be unstable.

\begin{figure}
\centerline{\includegraphics[width=.5\textwidth]{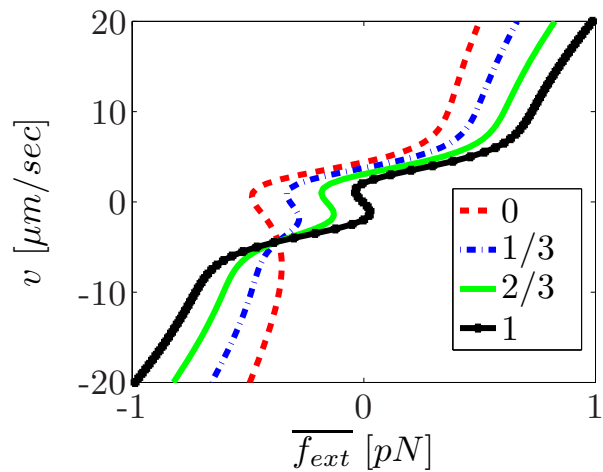}}
\caption{Results for different numbers of motors of each type. Shown is the evolution of the force-velocity curve as the ratio $N^{-}/N^{+}$ (given in the legend) changes. $\overline{f_{ext}}$ is normalized by $N^+$. Here $\kappa/k_a=5,~G=5 pN,~W=10 (\mu m/sec)^2$, $v_0=3 \mu m/sec$, $\gamma =1\mu m /sec$ and $\eta =0.02\mu m /sec$  }
\label{fig:FVevol} 
\end{figure}

We now turn to discuss the case where $N^+ \neq N^-$. The force-velocity curves are once again found using the antisymmetric combinations of the single force-velocity curves (see equation \eref{eq:2FVR}). When this is done  the force-velocity curve is no longer  antisymmetric. Changing the ratio $N^{+}/N^{-}$ leads to a continuous change of the force-velocity curve as shown in  \fref{fig:FVevol}. As expected, as the ratio of the number of motors increases the curves change continuously from the behavior of two sets of motors to that of a single set of motors. In general this leads to a loss of possible  bistable regions of the force-velocity curve.
 
Finally, we note that using the parameters from the fit of  \fref{fig:fit} it is possible to predict the force-velocity curve for two groups of myosin motors. We expect a force-velocity curve of type $(II)$, namely at zero force there are two possible velocities. Note, however, that care must be taken with this conclusion since we applied a specific (simple) model to fit the experimental data. For a clear conclusion the procedure explained above should be performed on data using constant velocity experiments (for example, using single molecule techniques).
\section{Dynamics\label{dynamics}}
We now discuss the dynamics of two sets of motors pulling oppositely in a tug-of-war. As above we assume a setup where $N^+$ motors pull against $N^-$ motors. We focus on the generic case (as we argue below) that $N^+\approx N^-$ and $\overline{f_{ext}}$ is small.   The more conventional setup of disordered motility assays will be discussed in the next section. 

The mean-field treatment of above does not account for the dynamics, and only provides possible steady state solutions. To account for the dynamics we expand the master equation of the process in powers of $1/N^{\pm}$ into a Fokker Planck equation. These results, and all other numerical results presented in the paper, are verified using Monte-Carlo simulations, described in \ref{appa}. 

As we show below two generic behaviors which depend on the presence of a bistability  in the mean-field solution are found. When there is no bistability (regimes (I) and (III)), as intuitively clear, the motion on long time-scales is a biased diffusion. The bias vanishes for $N^+=N^-$ at $\overline{f_{ext}}=0$, while for $N^+ \neq N^-$ the bias vanishes for a non zero value of $\overline{f_{ext}}$. Note however, that on short time scales the motion may display an oscillatory like behavior as seen in \fref{fig:anticor}(b). The velocity distribution function can not be fitted using a single Gaussian function (see in \fref{fig:anticor}(a)).

When a bistability is present (regimes (II) and (IV)) the behavior is more interesting. We find stochastic transitions between the two mean-field solutions. The motion on each of the two mean-field solutions is a biased diffusion with the bias dictated by the corresponding velocity. Each is  characterized by  an average dwell time, $\tau_+$ and $\tau_-$ for positive and negative velocities respectively. These are defined by the average time in which one solution changes to the other. Of course on long time scales the behavior is still a biased diffusion. The  analysis below shows that the average dwell times increase exponentially with the number of motors. The prefactor in the exponential depends on the exact value of the external force (This is very similar to the exponential time scales found for the symmetric motors in \cite{JulicherSim}). 
For $N^+ = N^-=N$ the ratio of the dwell times, $\tau_+/ \tau_-$, is exponentially large (small) in the number of motors for $\overline{f_{ext}}>0$ ($\overline{f_{ext}}<0$), namely $\tau_+/\tau_-\sim \exp(\theta N \overline{f_{ext}})$ where $\theta$ is a positive constant. This implies that for a large number of motors the transition between the two branches becomes very sharp. Specifically, in the $N\rightarrow \infty$ limit at $\overline{f_{ext}}=0$ the system spontaneously breaks between the two directions of motion. As discussed in \cite{JulicherSim} a similar picture also holds  for a single set of motors around the force regime exhibiting bistability. Note, that when no bistable region is present the corresponding dwell times of the oscillations like behavior (see \fref{fig:anticor}(b)) has a weak dependence on the number of motors. 

To derive these results we start by analyzing the case of a single set of motors which is then easily generalized to two set of motors.
Our model for one set of motors is a one step process with $N$ distinct states defined by the number of attached motors, $N_a$. 
The transition rate from a state with $N_{a}$ motors to one with  $N_a + 1$, is given by $g_{N_a}=k_a (N-N_a)$.  The transition from $ N_a$ to  $N_a -1 $ is given by $r_{N_a} = N_a k_d (v)$. The master equation then reads
\begin{equation}
\partial_t p(N_a) = -(r_{N_a}+g_{N_a})p(N_a) + r_{N_{a+1}} p(N_{a+1}) + g_{N_{a-1}} p(N_{a-1}),
\end{equation} 
  where $p(N_a)$ is the probability of having $N_a$ motors attached. Next, we define $q=N_{a}/N$ so that for every $N$, $q$ takes  values in the range $0$ to $1$. For brevity we use the operator $\mathbb{E}$, defined through its operation on a function $f(q)$, namely  $\mathbb{E}f(q)=f(q+1/N)$ and $\mathbb{E}^{-1} f(q)=f(q-1/N)$. It is easy to see that the master equation using these notations is given by
\begin{equation}
\partial_t p(q) = (\mathbb{E}-1)p(q)r(q)+(\mathbb{E}^{-1}-1)p(q)g(q),
\label{eq:Master}
\end{equation}
with $r(q)=r(N q)$ and $g(q)=g(N q)$. Next the operator $\mathbb{E}$ is expanded in powers of $1/N$ \cite{VanKampen} keeping terms up to second order so that: $\mathbb{E}=1+\frac{1}{N} \partial_q+\frac{1}{2N^2} \partial_{qq}$ and   $\mathbb{E}^{-1}=1-\frac{1}{N} \partial_q+\frac{1}{2N^2}  \partial_{qq}$. This gives the Fokker-Planck equation:
\begin{equation}
\partial_t p(q) = - \partial_q F_q p(q)+ \frac{1}{2N} \partial_{qq}  (D_q p(q)). \label{eq:Focker1}
\end{equation}
The right hand side of the equation has two terms: a drift term defined through,
\begin{equation}
F_{q} = \frac{1}{N} (g(q)-r(q)) = k_a (1-q)-q k_d(v),
\end{equation}
which for a constant $f_{ext}$ is independent of $N$. Note that requiring $F_q = 0$ yields the steady-state  mean-field equations. Therefore, the previous mean-field solutions serve as extrema  of the effective potential of the Fokker-Planck equation. It is easy to verify that solutions which satisfy $\partial_v f_{ext} < 0$ correspond to maxima while those with $\partial_v f_{ext} > 0$ correspond to minima. The second term is a diffusive term, with
\begin{equation}
D_q=\frac{1}{N} (g_{N q}+r_{N q})=k_a (1-q)+q k_d(v),
\end{equation}
independent of $N$ for a constant $f_{ext}$. In both the expressions of $F_q$ and $D_q$ the velocity is given by the force balance equation: $\eta  v = q  (G-\gamma v) +f_{ext}$. Note the overall $1/N$ term multiplying the diffusive term. A standard Kramers analysis implies that the time needed to cross the barrier from one minima to the other is given by $\tau \sim \exp(\alpha N)$ where $\alpha$ is a positive constant which depends on the functional form of $D_q$ and $F_q$. The analysis shows, similar to \cite{JulicherSim}, that the time increases exponentially with $N$.

Next we extend the above analysis to two sets of motors. We focus on the case $\overline{f_{ext}}=0$. Repeating the same procedure used to obtain  \eref{eq:Focker1} but now with the two coordinates $q=N^+_a/N^+$ and $s=N^-_a/N^-$, one obtains:
\begin{eqnarray}
\partial_{t}p & = & -\partial_{q}\left(F_{q}p\right)-\partial_{q}\left(F_{s}p\right)\nonumber \\
 &  & +\frac{1}{2N^{+}}\partial_{qq}\left(D_{q}p\right)+\frac{1}{2N^{-}}\partial_{ss}\left(D_{s}p\right).\label{eq:fokkerplanck}\end{eqnarray}
$F_q$ and $D_q$ are defined as above and $F_s$ and $D_s$ are given by:
\begin{eqnarray}
F_s & = & k_a (1-q)-q k_d(-v), \nonumber \\
D_s & = & k_a (1-q)+q k_d(-v).
\end{eqnarray}
On the right hand side of  \eref{eq:fokkerplanck} there are now two drift terms and two diffusive terms, one for each direction.  This equation is augmented by $ \eta (N^+ + N^-)=q N^+(G-\gamma v)- s N^- (G+\gamma v)$.  To analyze the equations we consider first the case $N^+=N^-$. In this case it easy to see that, due to the symmetry of the force balance equation, $F_q$, $F_s$, $D_q$ and $D_s$ become  independent of $N^{\pm}$. As in the case of a single a set of motors, the steady-state mean-field solution corresponds to $F_q=0$ and $F_s=0$. 

\begin{figure}
  \centering
  \subfloat[]{\label{fig:ForceField1}\includegraphics[width=.45\textwidth]{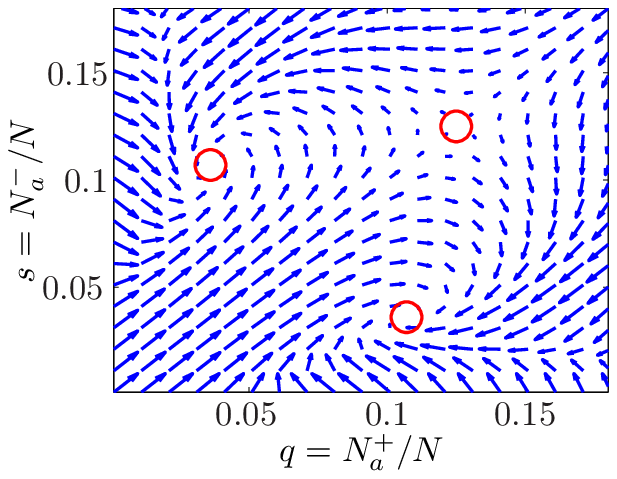}}
  \subfloat[]{\label{fig:bias1}\includegraphics[width=.45\textwidth]{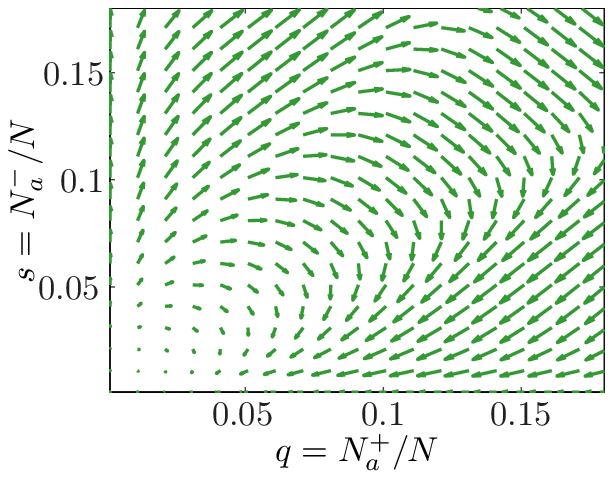}}
    \caption{The effective force-field  that governs the dynamics when there is a bistable region. (a) The effective force-field in the  $s$ , $q$ plane plotted when $N^+=N^-$. The three stationary points are marked by circles. Two of them are stable points while the third is a saddle point. (b) The additional contribution of the force-field that result for $N^+>N^-$. Here we set: $\kappa/k_a=5$, $W=10~ (\mu m /sec)^2$, $v_0=2~ \mu m /sec$, $G=5~ pN$, $\eta=0.02~ pN sec/\mu m$ and  $\gamma=0.4~ pN sec/\mu m$. }
  \label{fig:animals}
\end{figure}

The effective force-field may have either  one  or two stable stationary points that correspond to the number of the stable mean-field solutions (we ignore the rare possibility of four bistable regimes). We begin by discussing the latter and argue that it leads to ``oscillations'' between the two solutions. In   \fref{fig:ForceField1} the effective ``force-field''  caused by $F_q$ and $F_s$ for general values of $s$ and $q$ is shown. One can clearly see two stable minima and a single stationary unstable point (saddle point). The trajectories $\partial_t q=F_{q}(q,s)$ and $\partial_t s=F_{s}(q,s)$ are  exact in the limit $N^{\pm} \to \infty$. It is straightforward to argue that $\tau_+=\tau_-=\exp(\beta N^+)=\exp(\beta N^-)$. Here $\beta$ is a positive constant and $\tau_+$ and $\tau_-$ are the typical dwells time in the minima corresponding to the motors pulling to the right and left respectively. 
 
Generically $N^+$ and $N^-$  are not equal. In particular one expects for a typical bundle, composed of a random assortment of filaments with opposite polarity (see \fref{fig:setup1}), $N^+-N^- \sim \sqrt{N^+ + N^-}$. In this limit the mean-field solutions are modified from the case $N^+=N^-$ by terms of the order $1/\sqrt{N^+ + N^-}$, vanishing in the large $N^{\pm}$ limit. In contrast, we find that the dwell times are very sensitive to the difference in the number of motors of each type. In particular we find $\tau_+ / \tau_- \sim \exp( \beta N^+ -\beta N^-)$. Namely, when oscillations are present a small relative difference in the number of motors can lead to a strong asymmetry in the dwell times.

To see this dependence we consider the case $N^{\pm}=N (1 \pm \epsilon)$ with  $\epsilon$ of the order $1/\sqrt{N}$. The force-balance equation now gives $v$ an explicit dependence on the values of $N_+$ and $N_-$. This in turn leads to corrections of the order of $\epsilon$ to both $F_q$ and $F_s$, which are easily found to first order:
\begin{eqnarray}
\delta F_{q} &=&-\epsilon2q\frac{\kappa}{W}\left(v-v_{0}\right)\delta v, \nonumber\\
\delta F_{s}&=&-\epsilon2s\frac{\kappa}{W}\left(v+v_{0}\right)\delta v,\\
 \delta v &=& \left[\frac{Gq+Gs}{2\eta+\gamma q+\gamma s}-\frac{\left(q-s\right)^{2}G\gamma}{\left(2\eta+\gamma q+\gamma s\right)^{2}}\right]. \nonumber
\end{eqnarray} 

\begin{figure}
\centerline{\includegraphics[width=.5\textwidth]{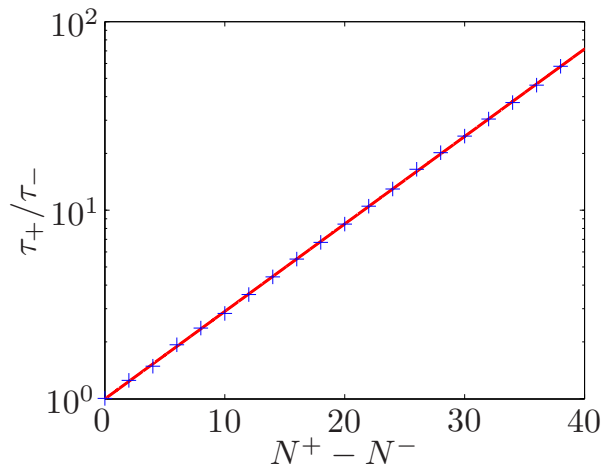}}
\caption{ The ratio of the dwell times, $\tau_+/\tau_-$, in a bistable region when $N^+ \neq N^-$ and there is no external force as obtained from numerical simulations. As can be seen to a good approximation  $\tau_+/\tau_- \propto \exp(\alpha N^+ - \alpha N^-)$, where $\alpha$ is a positive constant. Here $\kappa/k_a=1$, $W=10~ (\mu m /sec)^2$, $v_0=3~ \mu m /sec$, $G=5~ pN$, $\eta=0.0125~ pN sec/\mu m$ and  $\gamma=0.125~ pN sec/\mu m$.  }
\label{fig:Nassym}
\end{figure}

The contributions of these terms, as seen in  \fref{fig:bias1}, increase the transition probability from one minimum to the second and decrease the transition in the opposite direction (depending on the sign of $\epsilon$).
The analysis follows as above and we find $\tau_+\propto\exp(\alpha N + \kappa \epsilon N)$ and $\tau_-\propto\exp(\alpha N - \chi \epsilon N)$ with $\alpha,\kappa$ and $\chi$ positive constants. This implies that
\begin{equation}
\tau^+/\tau^- \sim \exp((\kappa+\chi)\epsilon N), 
\end{equation}
so that the ratio of the dwell times increases exponentially with the difference between the number of motors of each type. Note that the analysis relies on a relative small difference in the number of motors and in general holds as long as $\epsilon$ is small. When this is not the case the results might change. In particular, the mean-field solutions, as discussed in the previous section, might not show a bistable behavior. To verify the predicted behavior we have carried out numerical simulations. In  \fref{fig:Nassym} we plot the ratio $\tau^+/\tau_-$ and show that indeed it behaves as discussed above. 

\begin{figure}
\centerline{\includegraphics[width=.5\textwidth]{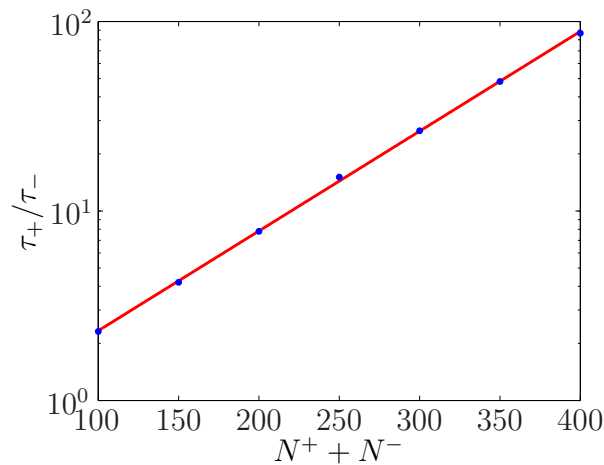}}
\caption{The ratio of dwell times, $\tau_+/\tau_-$, in a bistable region as a function of   $N^+ +  N^-$  for $N=N^+=N^-$ and a constant $\overline{f_{ext}}$ as obtained from numerical simulations. As shown to a good approximation,  $\tau_+/\tau_-\propto \exp(\theta \overline{f_{ext}} N)$. Here $k_a/\kappa=1$, $W=10~ (\mu m /sec)^2$, $v_0=2.5~ \mu m /sec$, $G=5~ pN$, $\eta=0.01~ pN sec/\mu m$, $\gamma=0.5~ pN sec/\mu m$ and $\overline{f_{ext}}=0.12~ pN$. }
\label{fig:tfN}
\end{figure}

\begin{figure}
\centerline{\includegraphics[width=.5\textwidth]{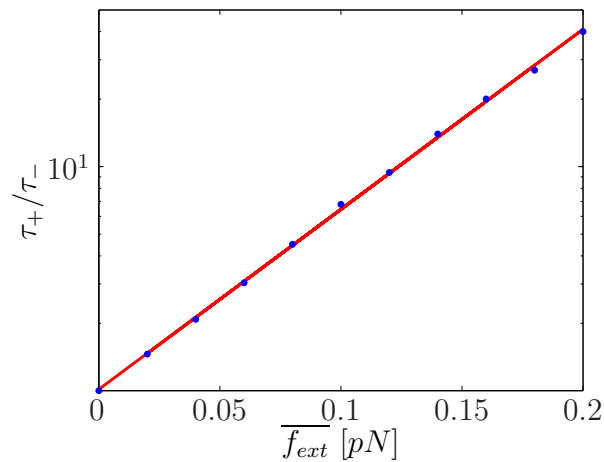}}
\caption{The ratio of dwell times in a bistable region for a constant  $N=N^+ =  N^-$ as a function of  $\overline{f_{ext}}$ as obtained from numerical simulations. As can be seen, to a good approximation, $\tau_+/\tau_-\propto \exp(\theta \overline{f_{ext}} N)$. All the parameters are the same as in  \fref{fig:tfN} except for $v_0=3 \mu m/sec$.}
\label{fig:tf}
\end{figure}

A similar effect occurs when $N^+=N^-=N$ and an external force is added. To first order in $\overline{f_{ext}}$ the effective forces $F_q(s,q)$ and $F_s(s,q)$ have the additional terms:
\begin{eqnarray}
\delta F_{q}=-2q\frac{\kappa}{W}\left(v-v_{0}\right)\frac{\overline{f_{ext}}}{2\eta+\gamma s+\gamma q},\nonumber\\
\delta F_{s}=-2s\frac{\kappa}{W}\left(v+v_{0}\right)\frac{\overline{f_{ext}}}{2\eta+\gamma s+\gamma q}.\\
\end{eqnarray}
It is straightforward to argue that $\tau_+/\tau_- \propto \exp( \theta N \overline{f_{ext}})$, where $\theta$ is positive constant. This is seen in figures \ref{fig:tfN} and \ref{fig:tf}. Note that this implies that for large $N^{\pm}$ the velocity, $v(\overline{f_{ext}})$, switches sharply between the positive and negative mean-field branches (see \fref{fig:fit}). In the limit $N \rightarrow \infty$ at $\overline{f_{ext}}=0$ the system exhibits spontaneous breaking of the symmetry between the two directions of motion. 


\begin{figure}
\centerline{\includegraphics[width=.5\textwidth]{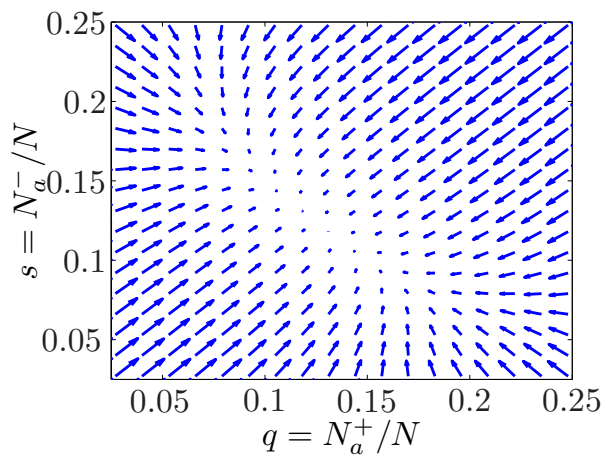}}
\caption{ The effective force-field when there is a single steady-state mean-field solution plotted in the $q$, $s$ plane. Note that there is only one stationary stable point. The shape of the effective force-field induces anticorrelation between $q$ and $s$. The parameters are the same as in  \fref{fig:ForceField1} except for $\eta=0.01~ pN sec/\mu m$ and  $\gamma=2.5~ pN sec/\mu m$. }
\label {fig:ForceField2}
\end{figure}

\begin{figure}
\centerline{\includegraphics[width=.6\textwidth]{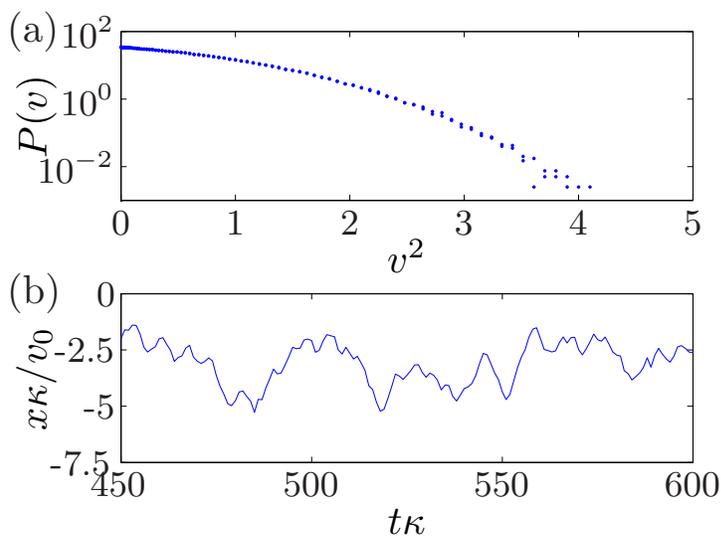}}
\caption{Dynamics in the absence of a bistability that  result in an oscillatory like motion. (a) The velocity probability distribution obtained from numerical simulations. (b) A representative trajectory  $x(t)$ showing an oscillatory like motion. The parameters used are the same as in  \fref{fig:ForceField2}. }
\label {fig:anticor}
\end{figure}

We now turn to discuss the case where there is no bistable region and $N^+ = N^-$. Since there is only one steady-state mean-field solution, the effective force-field has only one stationary point. Clearly, the motion on long time scales is therefore diffusive. However, on short time scales the motion may display a weak oscillatory like behavior.  \Fref{fig:ForceField2} shows an effective force that leads to this behavior. At the vicinity of the stationary point the force on the constant $q+s$  line is smaller than the force on the $q=s$ direction. This induces  anticorrelations between $q$ and $s$ which increases the probability to find a non zero velocity.  Typical traces of the displacement of the bundle, $x(t)$, appear almost bidirectional and the velocity distribution deviates from a Gaussian distribution as seen in  \fref{fig:anticor}(a). Note that for other choices of parameters, $q$ and $s$ can be positively  correlated, yielding a Gaussian distribution. Finally, as clearly evident in the case $N^+ \neq N^-$ the average velocity is no longer zero. 

In summary, irrespective of the presence of the bistability in the mean-field solution for a fixed $N^+$  and $N^-$, the motion on long-times and large-length scales is a biased diffusion. When a bistability is present the motion will exhibit oscillations in the sense described above and the crossover to the final biased diffusive behavior is expected to occur on time scales which are exponentially large in the number of motors. 

\section{Disorder\label{disorder}}
So far the number of motors pulling in each direction $N^{\pm}$ was taken to be a constant independent of the displacement of the filament. In typical motility assays the filament is moved around by  motors which  are bound to a substrate. Since the motor distribution is expected to be random this implies that in general $N^\pm$ depends on the location of the center of mass of the filament, its orientation and possibly time. In a typical experiment  one would expect  $N^{\pm}$ to be a random variable. This results, for example, from, as stated above, inhomogeneities of the motor density, randomly orientated motors and the motion of the motor head. In our analysis we ignore orientational changes of the bundle and assume that the motion is along a single axis. In motility assay experiments one dimensional motion is expected to occur on short time scales or by restricting the motion by methods used in \cite{Electric}. In fact some experiments show a motion very close to one dimensional \cite{AnnBiDirectional}.

\noindent Two extreme cases may be considered:\\
 \noindent \textit{(a) $N^{\pm}$ is only time dependent}. This could result, for example, from extremely flexible motors and  a bundle composed of a fine mesh of filaments. This allows each motor to bind to a positive and negative filament with equal probability. The motion  in this case on long-time scales and large-length scales will clearly be biased diffusion with the average velocity determined by the average of $N^+ -N^-$. A non-zero average $N^+ -N^-$ could result, for example, from a difference in the amount of filaments of each polarity which compose a bundle. To conclude, in this case the additional randomness does not change the qualitative motion however it could change the diffusion constant and the bias.\\
    \noindent \textit{(b) $N^{\pm}$ is only $x$ dependent}. Here  $x$ denotes the location of the center of mass of the bundle. This can occur, for example, if each motor can  bind only to one of the directions - either the plus oriented filaments or the negative oriented filaments and will naturally be the case when the bundle of the filaments in each direction is so thick that a motor can bind only to a filament of a given direction (or if the flexibility of the tail is limited).
    This coupled to inhomogeneities of the motor density is likely to lead to an $x$ dependent $N^{\pm}$ (see \fref{fig:setup1} where  $N^\pm$ is the integral of the motor density beneath the plus and minus filaments respectively).        
      In reality it is probable that the situation is a mix with both a time and an $x$ dependence with sensitivity to the  details of both the structure of the bundles and the elasticity of the motor tail. In such a case, as evident below, the $x$ dependence will dominate. 

\begin{figure}
\centerline{\includegraphics[width=.5\textwidth]{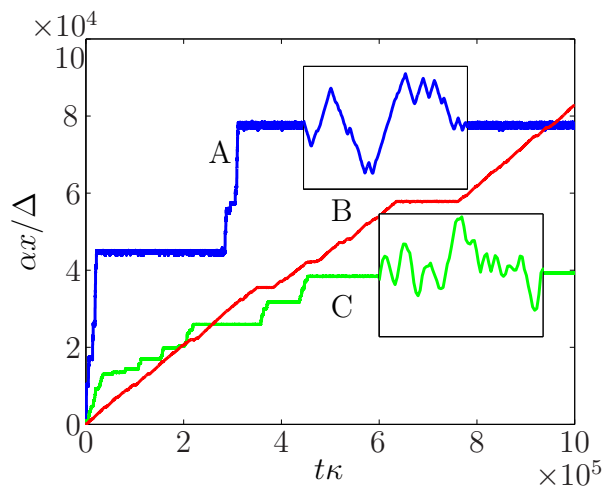}}
\caption{ Representative,  $x(t)$, trajectories obtained from simulations with different values of $\mu$. $\Delta$ is the distance between adjacent motors and $\alpha$ is a scale factor which allows us to show all three curves clearly on the same graph. In blue (A) a trajectory when there is a bistability, with  $\mu \simeq 0.4$ and $\alpha=1$.  In red (B) there is  a bistability , with $\mu  \simeq 1$ and $\alpha=0.077$. In green (C) there is no bistable region, with  $\mu \simeq 0.6$ and $\alpha=0.008$. Inside boxes is  a zoom on the motion. Parameters are the same as the ones in figures \ref{fig:muO} and \ref{fig:muNO}.}
\label{fig:typxt1}
\end{figure}
When $N^{\pm}(x)$ has an explicit  $x$ dependence the motion is more interesting. The  bundle appears to be trapped for long times at certain locations (see \fref{fig:typxt1} for a sample trajectory obtained from numerics). When this happens the motion of the bundle appears oscillatory like even if there is no bistability (see \fref{fig:typxt1}, green - C). 

Specifically $N^+(x)$ and $N^-(x)$ are now random variables with averages $\overline{N^+(x)}$ and  $\overline{N^-(x)}$ respectively. The overline denotes an average over $x$ locations (or equivalently an average over disorder realizations). In actual experiments it is more likely that a bundle composed of many randomly oriented filaments will not be symmetrical, namely $\overline{N^+(x)} \neq \overline{N^-(x)}$. Therefore $\overline{N^+(x)-N^-(x)}$, which signifies the overall polarity of the bundle, is expected from the central limit theorem to scale as $\sqrt{\ell}$, where $\ell$ is overall length of the bundle. 

When $N^+ \neq N^-$, as discussed above, there is a non-zero mean velocity implying that $N^+ - N^-$ acts as an effective force. The random $N^+(x)-N^-(x)$ thus induces a random effective force-field for the motion of the center of mass of the bundle. 

We expect for such problems, that on large-length scales and long-times, the systems can be described by a particle diffusing in an effective random forcing field. This problem has been studied extensively in the past \cite{AnomDiff} and one finds four prominent types of behaviors:\\
$(1)$ $\mu >2$ ,$ \overline {\langle x(t)\rangle}\sim t$ and $\overline {\langle x(t)^2 \rangle} -\overline {\langle x(t)\rangle}^2\sim t$. \\
(2)  $1<\mu<2$ - $\overline {\langle x(t)\rangle} \sim t$ and $\overline {\langle x(t)^2 \rangle} -\overline {\langle x(t)\rangle}^2 \sim t^{2/ \mu}$. \\
$(3)$ $0<\mu<1$  - $\overline {\langle x(t)\rangle} \sim t^{\mu}$ and $\overline {\langle x(t^2) \rangle} -\overline {\langle x(t)\rangle}^2 \sim t^{2 \mu}$.\\
$(4)$  $\mu=0$ - Sinai diffusion - $F_0 =0$,  $\overline{\langle x(t) \rangle}=0$  and $\overline{\langle x(t)^2 \rangle } \sim log^4(t/t_1)$.\\
The exponent $\mu=2 F_0 D/ \sigma$ is related to $F_0$, the average force acting on the particle, $D$, the diffusion constant in the absence of disorder and $\sigma$ is defined through the correlation $\overline {F(x)F(x')}  = \sigma \delta(x-x')$. The overline denotes an average over disorder realizations, angular brackets denote an average over histories of the system with a given realization of disorder and $f(x)=F(x)+F_0$ is the force acting on the particle at $x$ with the choice $\overline{F}=0$. Anomalous dynamics occur for $\mu<2$ and more prominently for $\mu<1$, and result from particles being trapped in rare deep wells, created by the random forcing field, for exponentially long times in the depth of the well \cite{AnomDiff}. 

\begin{figure}
\centerline{\includegraphics[width=.5\textwidth]{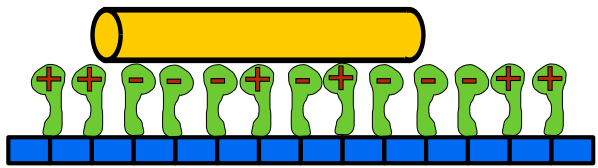}}
\caption{The model we use to  capture the effects of  disorder. The bundle (yellow) is set on a surface of random motors with either a plus or minus pulling direction. In case (a) the pulling direction at each site changes randomly with time. In case (b) the pulling direction is chosen randomly and does not change with time.}
\label {fig:Dismodel}
\end{figure}
To check if this straightforward analogy holds we account for disorder by defining a model, illustrated in  \fref{fig:Dismodel},  where $N^{\pm}(x)$ are random variables that depend on $x$, the location of the center of mass of the bundle. To generate a random motor landscape  the motors are placed equally spaced, separated by the distance $\Delta$, on a one dimensional lattice and assigned a random pulling direction that does not change with time. We consider a bundle composed of two groups of filaments with opposite polarities. The plus filament has a length of $L_+$ and the minus filament $L_-$. Therefore the probability that a motor has a plus / minus pulling direction  is $p_{\pm} = L_{\pm}/\ell$, where $\ell = L_++L_-$. It is easy to verify that the average polarity is $\overline{N^+(x)-N^-(x)} \propto L^+-L^-$.  Note that since the motors are equally spaced the bundle is always subject to a constant number of motors  $N^+(x)+ N^-(x)$ proportional to $\ell$.

The scale of the effective random forcing energy landscape is set by the distance between adjacent motors $\Delta$. The resulting effective potential is therefore expected to be proportional to  $\Delta$. In simulations $\mu$ can be adjusted by varying $\Delta$ while keeping all other parameters constant. This, however, may be difficult to achieve in experiments since $\Delta$  is controlled by the motor density. To keep all other parameters constant the size of the bundle and the viscosity  must therefore also be scaled. Nonetheless the motor density is related to  $\mu$.

We now turn to consider how a typical bundle will behave. A typical bundle composed of many filaments is expected to have an average polarity $\overline{N^+(x)-N^-(x)} \propto \sqrt{\ell}$. For this reason we set $\overline{N^+(x)-N^-(x)} \propto \sqrt{N^+ + N^-}$.

\begin{figure}
\centerline{\includegraphics[width=.5\textwidth]{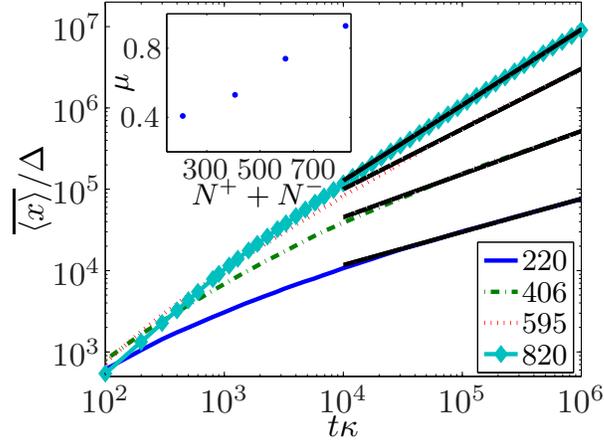}}
\caption{Disorder averaged trajectories, $\overline{\langle x(t) \rangle}$, in the presence of a bistability.  The black lines are the linear fits to the data which were used to obtain $\mu$. $\Delta=0.25 \mu m$ is the spacing between adjacent motors. In the inset $\mu$ is shown as a function of $N^++N^-$. Here  $\kappa/k_a=2$, $W=10~ (\mu m /sec)^2$, $v_0=2.5~ \mu m /sec$, $G=5~ pN$, $\eta=0.0125~ pN sec/\mu m$ and  $\gamma=1~ pN sec/\mu m$.}
\label {fig:muO}
\end{figure}

\begin{figure}
\centerline{\includegraphics[width=.5\textwidth]{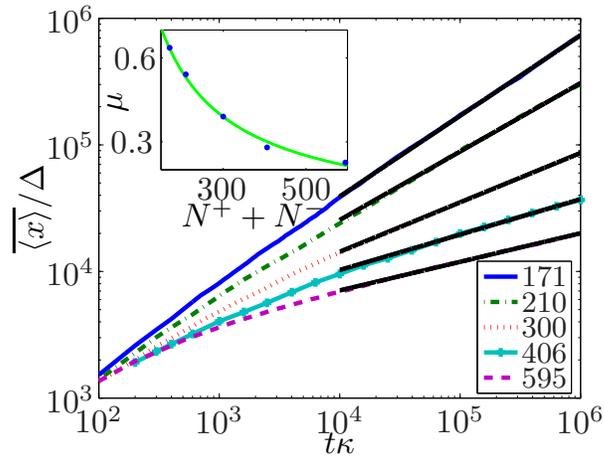}}
\caption{Disorder averaged trajectories, $\overline{\langle x(t) \rangle}$, when there is no bistability. The black lines are the linear fits to the data  used to obtain $\mu$. $\Delta=0.025 \mu m$ is the spacing between adjacent motors. In the inset $\mu$ is shown as a function of $N^++N^-$ and the line is a guide to the eye. $\kappa/k_a=2$, $W=10~ (\mu m /sec)^2$, $v_0=1~ \mu m /sec$, $G=5~ pN$, $\eta=0.0125~ pN sec/\mu m$ and  $\gamma=1~ pN sec/\mu m$. Parameters are the same as in \fref{fig:muO} except for $v_0=1~ \mu m /sec$.}
\label {fig:muNO}
\end{figure}

Using Monte-Carlo simulations and averaging over many realizations of disorder we verify that indeed the behavior is similar to the motion of a particle in a random forcing energy landscape and we are able to extract the exponent $\mu$ for different parameters from data over a few decades. We are interested in the dependence  of  $\mu$ on the bundle lengths $\ell$ in the bistable and non bistable regime. The results are given in  \fref{fig:muO} and \fref{fig:muNO}. We find two different behaviors for the bistable and non bistable regimes. For the  bistable regime $\mu$ grows with $N^++N^-$. Therefore in a motility assay experiment we expect that larger bundles will have larger $\mu$ (see \fref{fig:muO}). On long-time scales and large-length scales the motion will be a biased diffusion. On the other hand in the non bistable regime, as shown in \fref{fig:muNO},  $\mu$ decreases as $N^++N^-$ increases. Here the $x(t)$ grows slower for larger bundles and in the limit of an infinite bundle  $\overline{\langle x(t) \rangle} \rightarrow 0$ and $\overline{\langle x^2(t) \rangle}\sim log^4(t/t_1)$, with $t_1$ a time scale.

One characteristic of the motion in a random-forcing energy landscape with $\mu <1$ is the long period of time when the bundle is trapped in a ``deep potential well'', localized around a certain point. These are then  followed by a quick transition to another potential well (see \fref{fig:typxt1}). On average the trapping time increases with the observation time. The farther the particle traverses the more likely  it is to find a deeper well. Inside such a deep  well the bundle is trapped and its motion is oscillatory like even if there is no bistability. Even for $\mu>1$ on short time scales the motion still includes long trapping times where the bundle is stuck at a certain point.

We note that while in a tug-of-war where anomalous diffusion occurs naturally with no additional external force, it is possible to view the same effects by applying an external force to \textit{one set of motors}. The force must then be tuned to reach the region of average  zero  velocity.
\section{Conclusions\label{conclusions}}
The paper focused on the dynamics of motors in a tug-of-war situation which are coupled by a rigid backbone. Using a steady-state mean-field solution of a model introduced in \cite{vilfan} we characterized different possible force-velocity relations which in many cases exhibit regions of bistability. The implications for motility assays were then discussed. 

It is interesting to compare our results to recent motility assays on myosin motors with a setup very similar to the one we consider \cite{AnnBiDirectional}. In the experiments bundles of actin indeed exhibit trapping for long period of times in specific locations. When trapped they exhibit an oscillating like behavior between two velocities. This agrees well with our prediction for a random forcing energy landscape. 

\begin{figure}
\centerline{\includegraphics[width=.5\textwidth]{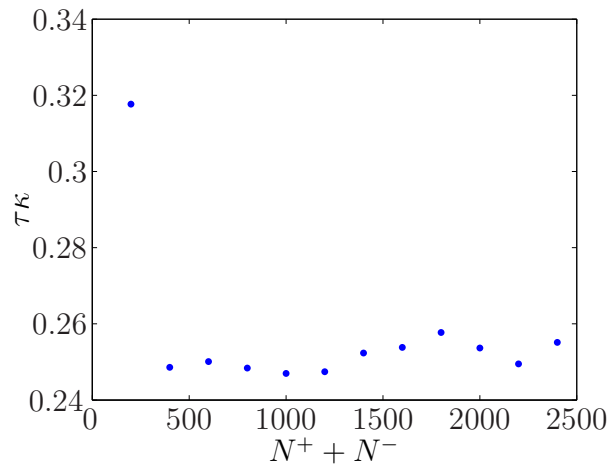}}
\caption{The dwell time, $\tau$, when there is no bistable region  and $\overline{N^+-N^-}=0$. The dwell time is obtained from a single realization of the disorder using methods described in \ref{appb}. As can be seen the dwell time has a weak dependence on $N^++N^-$. $x(t)$ is sampled in steps of $dt=0.01/ \kappa$. Here $\kappa/k_a=1$, $W=10~ (\mu m /sec)^2$, $v_0=2~ \mu m /sec$, $G=5~ pN$, $\eta=0.1~ pN sec/\mu m$ and  $\gamma=1.4~ pN sec/\mu m$. }
\label {fig:TNO}
\end{figure}

\begin{figure}
\centerline{\includegraphics[width=.5\textwidth]{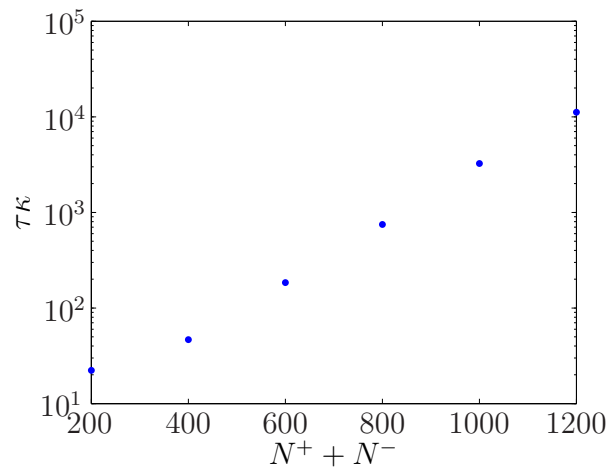}}
\caption{ The dwell time, $\tau$, when there is a bistable region  and $\overline{N^+-N^-}=0$. The dwell time is obtained from a single realization of disorder using methods described in \ref{appb}. As can be seen the dwell time, to a good approximation, grows exponentially with $N^++N^-$. The parameters are the same as in \fref{fig:TNO} except for $\gamma=0.1~ pN sec/\mu m$.}
\label {fig:TO}
\end{figure}

Furthermore, in \cite{AnnBiDirectional} they studied bundles trapped in such minima and considered the dependence of the average dwell time in each velocity as a function of the number of motors. The experiments show a weak dependence on the number of motors, which in \cite{AnnBiDirectional} was explained by cooperative unbinding of the motors. As shown in  \fref{fig:typxt1}, in a minima of the random forcing energy landscape, as intuitively clear, the motion of the bundle exhibits an oscillating like behavior. Moreover, when the force-velocity relation exhibits no bistability the dwell time in each velocity has a very weak dependence on the number of motors (see \fref{fig:TNO}). This provides an alternative explanation of the experimental results. This conclusion could possibly be verified by single molecule experiments of the force-velocity curve of a single set of motors under the experimental settings of \cite{AnnBiDirectional}. (We note that our analysis of \fref{fig:fit} suggests a bistable behavior for myosin. However, as stated above, care has to be taken since a specific model was assumed in fitting the experimental data of \cite{Electric} and the exact setup in \cite{Electric} and \cite{AnnBiDirectional} is somewhat different.) For a clear analysis constant velocity experiments have to be carried out followed by the analysis we describe. In \fref{fig:TO} we show the average dwell time when a bistability is present in the force-velocity curve. There the behavior is distinct and the dwell time in each velocity has a strong exponential dependence on the number of motors. Note that the dwell time in figures \ref{fig:TNO} and \ref{fig:TO} in principle depends on the realization of disorder and the local structure of the effective potential well.

Finally, we note that using the techniques described above it is straight forward to derive the behavior of motors of different type each pulling in an opposite direction.

\section*{Acknowledgment}

This work was supported by the Israeli Science Foundation. We are grateful for discussions with Anne Bernheim, Oded Fargo and  Frank  J\"{u}licher. We also thank Kinnert Kernen for a critical reading of the manuscript. 

\appendix

\section{Simulations\label{appa}}
Monte Carlo simulations are employed to verify mean-field results and  test our arguments. We use a standard Gillspie \cite{Gillespie} algorithm. In each time step one transition  occurs, namely one of $N^+ +N^-$ motors will either detach or  attach. Since  transition times are Poissonian the probability for each transition is proportional to its respective rate.  Next, the advancement time is chosen from an exponential distribution, with a time constant $1/T = N^+_a k_d(v) +N^-_a k_d(-v) +(N^+-N^+_a) k_a +(N^--N^-_a) k_a$. Once these two steps are completed the velocity and the rates are recalculated and the whole cycle is repeated.

In the case of disorder the definition of the rates and the algorithm are mostly unchanged. The only difference is that one needs  to account for the changes in $N^{\pm}(x)$. This is done by modifying only the number of detached motors $N^{\pm}_d(x)$ as the bundle moves along the motor landscape. The results are not expected to change due to this approximation even under a more detailed model as long as most of the motors remain in the detached state.

Throughout the paper $x(t)$ is sampled at time steps of $dt=100/\kappa$ and the length of the simulations is $10^6 /\kappa$ unless specified otherwise. 
\section{Dwell times measurements \label{appb}}

In several parts of the paper  the dwell times, $\tau_+$ and $\tau_-$, are estimated in the bistable region from the numerical simulations. These are extracted  from the autocorrelation function of $N^+-N^-$  which  we observe to decay exponentially  as,  $\exp(-t/T)$ with $T$ a time scale. In general $\tau_+ \neq \tau_-$  and  $1/T= 1/\tau_+ +1/ \tau_-$. To find $\tau_+$ and $\tau_-$ independently we calculate numerically the ratio of the time spent in the positive and in the negative velocity  which yields $\tau_+/\tau_-$. Then  together with the expression for $1/T$  $\tau_+$ and $\tau_-$ are obtained. 

Note that when there is disorder the autocorrelation tail may not be purely exponential because of the local  motor landscape. In this case the dwell time can be found directly from $x(t)$.  It is given by the average time it takes the velocity to switch signs. This method is useful when there is a single dwell time, namely  $\overline{f_{ext}}=0$ and  $\overline{N^+} = \overline{N^-}$.

\section*{References}

\providecommand{\newblock}{}

\end{document}